\documentclass{aastex6}

\begin{document}


\title{$^{13}$C Isotopic Fractionation of HC$_{3}$N in Star-Forming Regions \\ - Low-Mass Star Forming Region L1527 and High-Mass Star Forming Region G28.28-0.36 -}


\author{Kotomi Taniguchi\altaffilmark{1,2}, Masao Saito\altaffilmark{2}}
\affil{Department of Astronomical Science, School of Physical Science, SOKENDAI (The Graduate University for Advanced Studies), Osawa, Mitaka, Tokyo 181-8588, Japan}

\and

\author{Hiroyuki Ozeki}
\affil{Department of Environmental Science, Faculty of Science, Toho University, Miyama, Funabashi, Chiba 274-8510, Japan}


\altaffiltext{1}{kotomi.taniguchi@nao.ac.jp}
\altaffiltext{2}{Nobeyama Radio Observatory, National Astronomical Observatory of Japan, Minamimaki, Minamisaku, Nagano 384-1305, Japan}

\begin{abstract}

We observed the $J=9-8$ and $10-9$ rotational lines of three $^{13}$C isotopologues of HC$_{3}$N in L1527 and G28.28-0.36 with the 45-m radio telescope of the Nobeyama Radio Observatory in order to constrain the main formation mechanisms of HC$_{3}$N in each source.
The abundance ratios of the three $^{13}$C isotopologues of HC$_{3}$N are found to be 0.9 ($\pm0.2$) : 1.00 : 1.29 ($\pm0.19$) ($1\sigma$) and 1.0 ($\pm 0.2$) : 1.00 : 1.47 ($\pm 0.17$) ($1\sigma$) for $[$H$^{13}$CCCN$]: [$HC$^{13}$CCN$]: [$HCC$^{13}$CN] in L1527 and G28.28-0.36, respectively.
We recognize a similar $^{13}$C isotopic fractionation pattern that the abundances of H$^{13}$CCCN and HC$^{13}$CCN are comparable, and HCC$^{13}$CN is more abundant than the others.
Based on the results, we discuss the main formation pathway of HC$_{3}$N.
The $^{13}$C isotopic fractionation pattern derived from our observations can be explained by the neutral-neutral reaction between C$_{2}$H$_{2}$ and CN in both the low-mass (L1527) and high-mass (G28.28-0.36) star forming regions.

\end{abstract}




\section{Introduction} \label{sec:intro}

Almost 200 molecules have been detected in the interstellar medium or circumstellar shells so far.
In dark clouds, unsaturated carbon-chain molecules such as CCS are abundant, whereas they decrease in star forming cores \citep{1992ApJ...392...551}.
These carbon-chain molecules are formed from C or C$^{+}$ in the gas phase before carbon atoms are converted into CO.
On the other hand, saturated complex organic molecules such as CH$_{3}$OH and CH$_{3}$CN increase in star forming regions, and these regions are called hot cores for high-mass star forming regions and hot corinos for low-mass star forming regions.
In hot cores and hot corinos, the combination between grain-surface reactions and gas-phase reactions produces various saturated complex organic molecules \citep[e.g.][]{2006A&A...457...927}.

Recently, in contrast to the above scenarios, several star forming cores associated with carbon-chain molecules have been discovered.
\citet{2008ApJ...672...371} showed that various carbon-chain molecules are abundant toward the low-mass star forming region L1527.
Several chemical model calculations showed that CH$_{4}$ plays an essential role in efficient formation of carbon-chain molecules in the warm regions ($\geq25$ K) \citep[e.g.][]{2008ApJ...674...993, 2008ApJ...681...1385}.
\citet{2008ApJ...674...993} indicated that carbon-chain molecules are formed by a combination of gas-phase reactions and grain-surface reactions following the sublimation of CH$_{4}$.
These low-mass star forming regions were named Warm Carbon Chain Chemistry (WCCC) sources.

Cyanopolyynes (HC$_{2n+1}$N) are one of the representative carbon-chain-molecule series.
\citet{2008ApJ...681...1385} demonstrated that regeneration of carbon-chain molecules results from gas-phase chemistry and showed the primary formation pathway of HC$_{3}$N as follows:

\begin{equation} \label{r6}
{\mathrm {CN}} + {\mathrm C}_{2}{\mathrm H}_{2} \rightarrow {\mathrm {HC}}_{3}{\mathrm N} + {\mathrm H}.
\end{equation}

\citet{2009MNRAS...394...221} presented their time-dependent gas-phase chemical model and showed that cyanopolyynes (HC$_{2n+1}$N; $n=1-4$) can be formed under hot core conditions.
They considered that C$_{2}$H$_{2}$ is released from the grain mantle inside hot cores and the neutral-neutral reaction between C$_{2}$H$_{2}$ and CN (Reaction (\ref{r6})) proceeds relatively easily for the hot core temperature ($100-300$ K).
Their chemical model calculation was supported by detections of HC$_{5}$N toward 35 hot cores associated with 6.7 GHz methanol masers \citep{2014MNRAS...443...2252}.
Therefore, the predicted main formation pathway of HC$_{3}$N is Reaction (\ref{r6}) in both the low-mass star forming region L1527 and the high-mass star forming regions.

Deriving $^{13}$C isotopic fractionation of carbon-chain molecules by observations is one of the promising methods to constrain their main formation mechanism.
For example, the dominant formation mechanism of HC$_{3}$N has been investigated based on its $^{13}$C isotopic fractionation \citep{1998aap...329...1156} toward the cyanopolyyne peak in Taurus Molecular Cloud-1 (TMC-1 CP; $d=140$ pc).
The derived abundance ratios of three $^{13}$C isotopologues of HC$_{3}$N are $[$H$^{13}$CCCN$]: [$HC$^{13}$CCN$]: [$HCC$^{13}$CN] = $1.0:1.0:1.4$ ($\pm0.2$) ($1\sigma$), and it is suggested that its main formation pathway is the neutral-neutral reaction between C$_{2}$H$_{2}$ and CN \citep{1998aap...329...1156}.
Thus investigation of formation mechanism of HC$_{3}$N using its $^{13}$C isotopic fractionation has proved successful.

In the present paper, we observed the three $^{13}$C isotopologues of HC$_{3}$N, the shortest cyanopolyyne, toward L1527 ($d=140$ pc) and G28.28-0.36 ($d=3.0$ kpc, \citet{2014MNRAS...443...2252}) with the 45-m radio telescope of the Nobeyama Radio Observatory (NRO) in order to compare $^{13}$C isotopic fractionation of HC$_{3}$N and its main formation mechanisms in the two star forming regions.
G28.28-0.36 is one of the hot cores that \citet{2014MNRAS...443...2252} detected HC$_{5}$N, and associates with an ultra compact HII region.
\citet{2006MNRAS...367...553} detected CH$_{3}$CN, a hot core tracer, and H$^{13}$CO$^{+}$ toward the hot core.

\section{Observations} \label{sec:obs}

We carried out observations of the three $^{13}$C isotopologues of HC$_{3}$N (H$^{13}$CCCN, HC$^{13}$CCN, and HCC$^{13}$CN) simultaneously with the NRO 45-m radio telescope in 2015 December, 2016 Feburary, March and April (2015-2016 season).
We also observed the normal species of the $J=9-8$ rotational lines simultaneously, when we observed the $J=9-8$ lines of the three $^{13}$C isotopologues.
We used the TZ and the T70 receivers for the observations of the $J=9-8$ transition lines, and the TZ receiver for the observations of the $J=10-9$ transition lines.
We used the TZ receiver for the observations of the $J=9-8$ lines, after we could not use the T70 receiver due to equipment troubles.
The rest frequencies of the observed lines are given in Table \ref{tab:tab2}.
Both receivers allow us to obtain dual-polarization data simultaneously.
The beam sizes (HPBW) and main beam efficiencies ($\eta_{B}$) were 18'' and 54\% for the TZ receiver and 20'' and 56\% for the T70 receiver at 86 GHz, respectively.
The system temperatures were between 120 and 250 K, depending on the weather conditions and the elevations.
We employed the position-switching mode.

The observed position and off-source position for L1527 were ($\alpha_{2000}$, $\delta_{2000}$) = (04$^{\rm h}$39$^{\rm m}$53\fs89, 26\arcdeg03\arcmin11\farcs0) and (04$^{\rm h}$42$^{\rm m}$35\fs9, 25\arcdeg53\arcmin23\farcs3), respectively.
The telescope pointing was checked using the H40 receiver every 1.5 hours by observing the SiO maser line ($J=1-0$) from NML Tau, and the pointing error was less than 3''.
The observed position for G28.28-0.36 was ($\alpha_{2000}$, $\delta_{2000}$) = (18$^{\rm h}$44$^{\rm m}$13\fs3, -04\arcdeg18\arcmin03\farcs3), and the off-source position was set to be $+15$' away in the declination.
We checked the pointing accuracy by observing the SiO maser lines ($J=1-0$) from OH39.7+1.5 every $1-1.5$ hours, depending on the wind conditions.
The pointing error was within 3''.
We used the SAM45 FX-type digital correlator in frequency settings whose bandwidths and resolutions are 125 MHz and  30.52 kHz, and 250 MHz and 61.04 kHz for L1527 and G28.28-0.36, respectively.
We applied 2 channel binding, and velocity resolutions of final spectra are 0.25 and 0.5 km s$^{-1}$ for L1527 and G28.28-0.36, respectively.

\section{Results and Analysis} \label{sec:res}

\subsection{Results} \label{sec:res}

\begin{figure}[ht!]
\figurenum{1}
\plotone{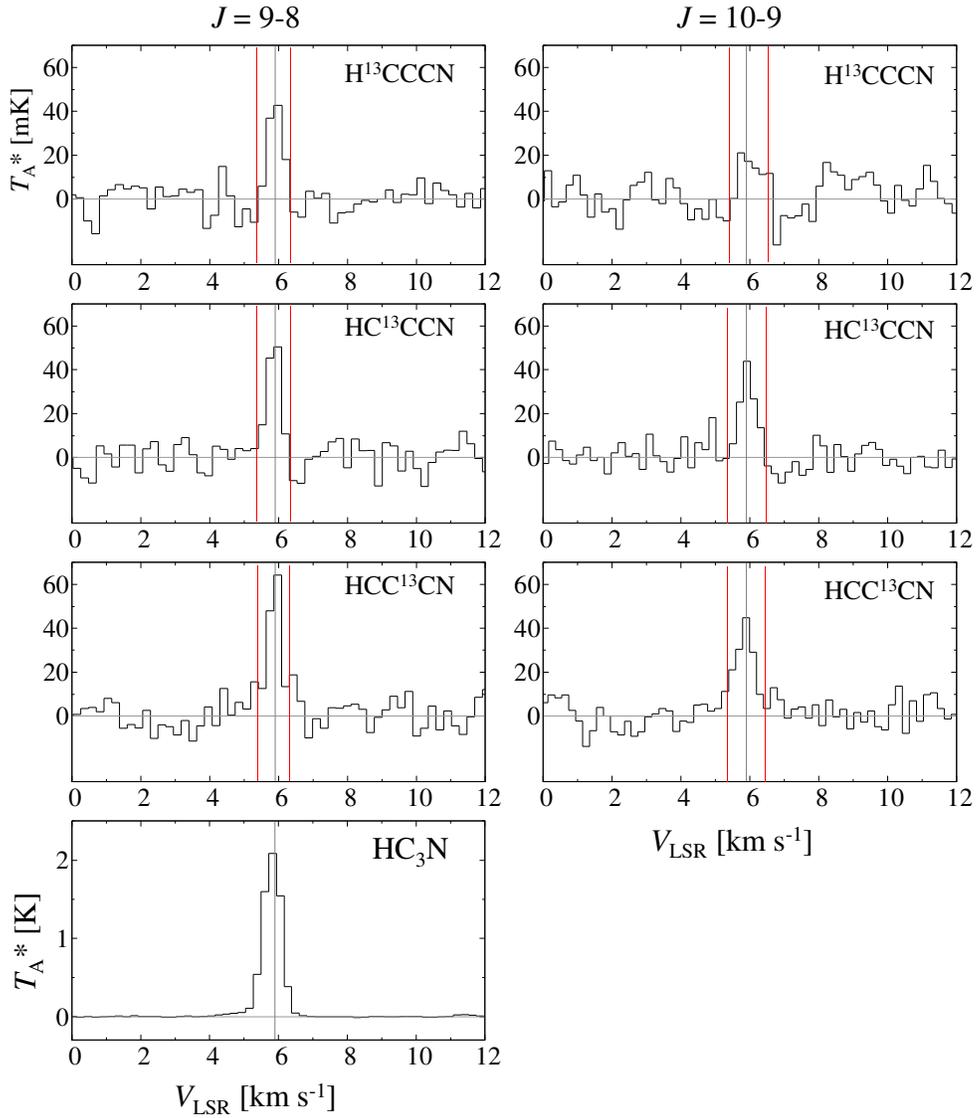}
\caption{The spectra of the three $^{13}$C isotopologues of HC$_{3}$N of the $J=9-8$ and $J=10-9$ rotational transitions and the normal species of the $J=9-8$ rotational transition toward L1527. The gray vertical lines show {\it V}$_{{\mathrm {LSR}}}=5.9$ km s$^{-1}$. The red vertical lines show the range using for the integrated intensities.\label{fig:f1}}
\end{figure}

The spectra of the three $^{13}$C isotopologues of HC$_{3}$N were taken with signal-to-noise ratios of $6.3-10.6$ in L1527, except the $J=10-9$ line of H$^{13}$CCCN, as shown in Figure \ref{fig:f1}.
We also show the spectra of the $J=9-8$ rotational line of the normal species observed with its $^{13}$C isotopologues simultaneously.
The values of {\it V}$_{\mathrm {LSR}}$ are in good agreement with the {\it V}$_{\mathrm {LSR}}$ value reported for this source (5.9 km s$^{-1}$).
We derived their integrated intensities with the same velocity range, and the results are summarized in Table \ref{tab:tab1}.
The ratios of the integrated intensities among the three $^{13}$C isotopologues of the $J=10-9$ lines are consistent with those of the $J=9-8$ lines.
However, we cannot detect H$^{13}$CCCN using the $J=10-9$ line with signal-to-noise ratio above 3, and we do not use the $J=10-9$ lines in the following analyses.
We evaluated the error of the integrated intensities using the following formula:

\begin{equation} \label{error}
\Delta T_{\rm {A}}^{*}\; ({\rm {K}}) \times \sqrt{n\; ({\rm {ch}})} \times v\; ({\rm {km\; s}}^{-1})
\end{equation}

In Equation (\ref{error}), $\Delta${\it T}$^{\ast}_{\mathrm A}$ is the rms noises in the emission-free regions, $n$ is the numbers of channels, and $v$ is the velocity resolution per channel.
The rms noises are summarized in Table \ref{tab:tab2}.
We used 4 ch and 0.25 km s$^{-1}$ for $n$ and $v$, respectively, in L1527.

\begin{figure}[ht!]
\figurenum{2}
\plotone{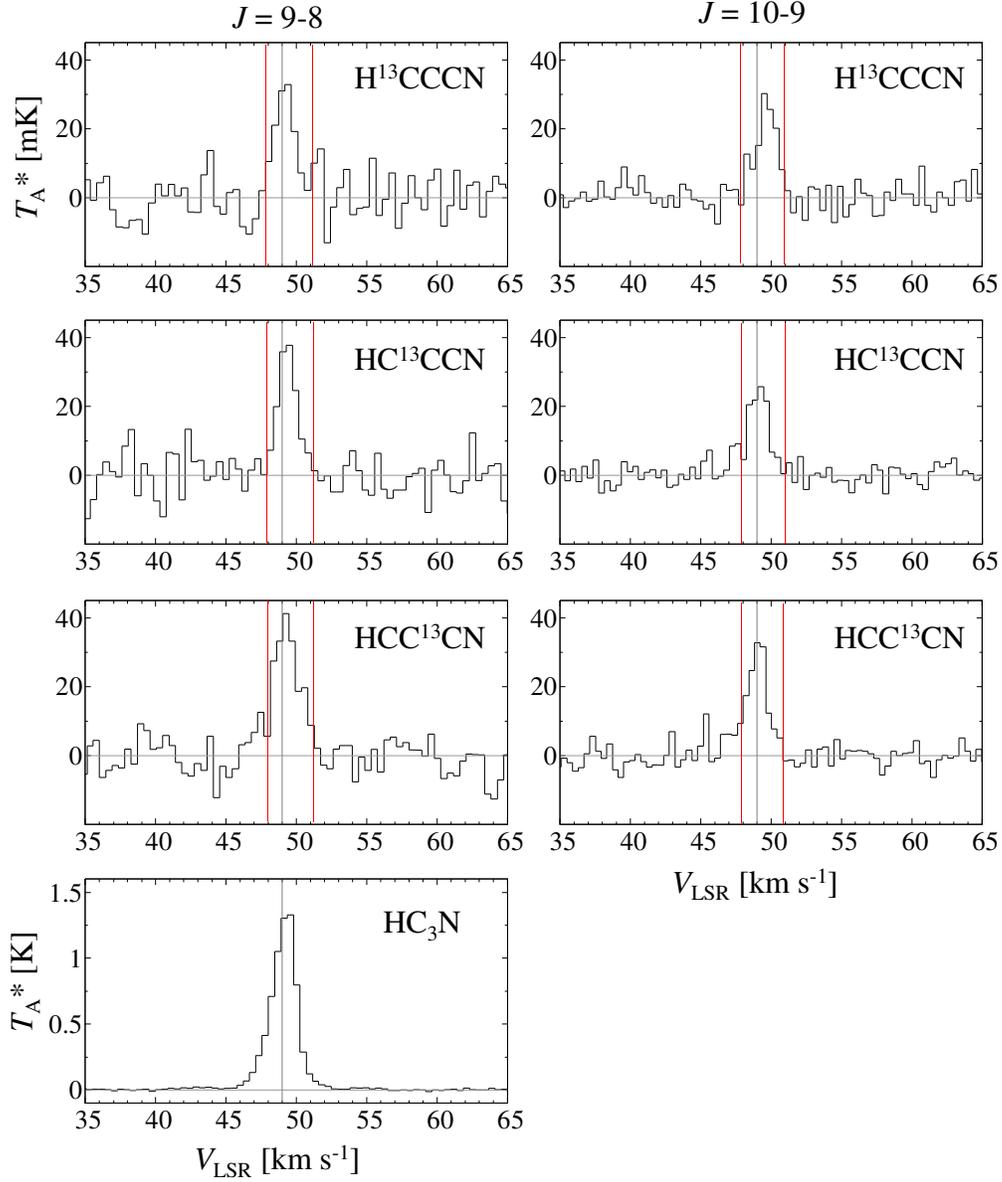}
\caption{The spectra of the three $^{13}$C isotopologues of HC$_{3}$N of the $J=9-8$ and $J=10-9$ rotational transitions and the normal species of the $J=9-8$ rotational transition toward G28.28-0.36. The gray vertical lines show {\it V}$_{{\mathrm {LSR}}}=48.9$ km s$^{-1}$.The red vertical lines show the range using for the integrated intensities.\label{fig:f2}}
\end{figure}

The three $^{13}$C isotopologues of HC$_{3}$N were detected with signal-to-noise ratios of $5.1-10.7$ in G28.28-0.36, as shown in Figure \ref{fig:f2}.
The values of {\it V}$_{{\mathrm {LSR}}}$ agree with one another and the systematic velocity reported for the source (48.9 km s$^{-1}$ \citet{2006MNRAS...367...553}).
We derived the integrated intensities of the three $^{13}$C isotopologues of HC$_{3}$N in the same way as L1527, and their errors were calculated using Equation (\ref{error}) with 7 ch and 0.5 km s$^{-1}$ for $n$ and $v$, respectively.
We summarize the results in Table \ref{tab:tab1}. 

 \floattable
\begin{deluxetable}{cccc}
\tabletypesize{\scriptsize}
\tablecaption{Integrated Intensities of Three $^{13}$C Isotopologues of HC$_{3}$N (K km s$^{-1}$)\label{tab:tab1}}
\tablewidth{0pt}
\tablehead{
\colhead{} & \colhead{H$^{13}$CCCN} & \colhead{HC$^{13}$CCN} & \colhead{HCC$^{13}$CN}
}
\startdata
L1527 & & & \\
$J=9-8$ & 0.021 (3) & 0.028 (3) & 0.035 (3) \\
$J=10-9$ & 0.016 (4) & 0.023 (3) & 0.032 (3) \\
 & & & \\
G28.28-0.36 & & & \\
$J=9-8$ & 0.055 (9) & 0.064 (7) & 0.090 (7) \\
$J=10-9$ & 0.049 (5) & 0.047 (4) & 0.059 (4) \\
\enddata
\tablecomments{The numbers in parenthesis represent the error values evaluated by Equation (\ref{error}).}
\end{deluxetable}


\floattable
\begin{deluxetable}{lccccc}
\tabletypesize{\scriptsize}
\tablecaption{Spectral Line Parameters of the Observed Lines\label{tab:tab2}}
\tablewidth{0pt}
\tablehead{
\colhead{Species} & \colhead{Frequency\tablenotemark{a}} & \colhead{{\it T}$^{\ast}_{\mathrm A}$} & \colhead{$\Delta v$} & \colhead{{\it V}$_{\mathrm {LSR}}$} & \colhead{rms\tablenotemark{b}} 
\\
\colhead{} & \colhead{(GHz)} & \colhead{(mK)} & \colhead{(km s$^{-1}$)} & \colhead{(km s$^{-1}$)} & \colhead{(mK)}
}
\startdata
\multicolumn{6}{l}{L1527} \\
\multicolumn{6}{l}{$J = 9-8$} \\
HC$_{3}$N & 81.8814677 (9) & 2154 (16)  & 0.622 (6) & 5.8 (3) & 5.5 \\
H$^{13}$CCCN & 79.35046 (2) & 45 (6) & 0.49 (7) & 6.0 (3) & 6.6 \\
HC$^{13}$CCN & 81.53411 (2) & 52 (5) & 0.46 (6) & 6.0 (3) & 5.9 \\ 
HCC$^{13}$CN & 81.54198 (2) & 66 (6) & 0.47 (5) & 6.0 (3) & 6.2 \\
\multicolumn{6}{l}{$J = 10-9$} \\
H$^{13}$CCCN & 88.16683 (2) & $<22$ & - & -  & 7.4 \\
HC$^{13}$CCN & 90.59306 (2) & 43 (5) & 0.51 (6) & 5.9 (3) & 5.9 \\
HCC$^{13}$CN & 90.60178 (2) & 39 (4) & 0.75 (9) & 5.9 (3) & 5.9 \\
\multicolumn{6}{l}{} \\
\multicolumn{6}{l}{G28.28-0.36} \\
\multicolumn{6}{l}{$J = 9-8$} \\
HC$_{3}$N & 81.8814677 (9) & 1310 (20) & 2.19 (7) & 49.2 (5) & 6.7 \\
H$^{13}$CCCN & 79.35046 (2)  & 34 (4) & 1.6 (2) & 49.4 (5) & 6.6 \\
HC$^{13}$CCN & 81.53411 (2) & 39 (4) & 1.6 (2) & 49.5 (5) & 5.3 \\ 
HCC$^{13}$CN & 81.54198 (2) & 38 (3) & 2.3 (2) & 49.3 (5) & 5.3 \\
\multicolumn{6}{l}{$J = 10-9$} \\
H$^{13}$CCCN & 88.16683 (2) & 28 (2) & 1.64 (17) & 49.5 (5) & 3.9 \\
HC$^{13}$CCN & 90.59306 (2) & 25 (2) & 1.86 (17) & 49.3 (5) & 2.9 \\                   
HCC$^{13}$CN & 90.60178 (2) & 31 (2) & 1.79 (15) & 49.0 (5) & 2.9 \\         
\enddata
\tablecomments{The numbers in parenthesis represent one standard deviation in the Gaussian fit except for frequency.}
\tablenotetext{a}{Taken from the Cologne Database for Molecular Spectroscopy (CDMS) \citep{2005JMoSt...742...215}.}
\tablenotetext{b}{The rms noises in emission-free regions.}
\end{deluxetable}


\subsection{Analysis}

We fitted the spectra with a Gaussian profile and obtained the spectral line parameters, as summarized in Table \ref{tab:tab2}.
We calculated the column densities of the normal species and the three $^{13}$C isotopologues of HC$_{3}$N using the Local Thermodynamic Equilibrium (LTE) analysis as shown in the following formulae \citep{1998aap...329...1156}:

\begin{equation} \label{tau}
\tau = - {\mathrm {ln}} \left[1- \frac{T^{\ast}_{\rm A} }{f\eta_{\rm B} \left\{J(T_{\rm {ex}}) - J(T_{\rm {bg}}) \right\}} \right],  
\end{equation}
where
\begin{equation} \label{tem}
J(T) = \frac{h\nu}{k}\Bigl\{\exp\Bigl(\frac{h\nu}{kT}\Bigr) -1\Bigr\} ^{-1},
\end{equation}  
and
\begin{equation} \label{col}
N = \tau \frac{3h\Delta v}{8\pi ^3}\sqrt{\frac{\pi}{4\mathrm {ln}2}}Q\frac{1}{\mu ^2}\frac{1}{J_{\rm {lower}}+1}\exp\Bigl(\frac{E_{\rm {lower}}}{kT_{\rm {ex}}}\Bigr)\Bigl\{1-\exp\Bigl(-\frac{h\nu }{kT_{\rm {ex}}}\Bigr)\Bigr\} ^{-1}.
\end{equation} 

In Equation (\ref{tau}), {\it T}$^{\ast}_{\mathrm A}$ denotes the antenna temperature, {\it f} the beam filling factor, $\eta_{\rm{B}}$ the main beam efficiency, and $\tau$ the optical depth.
We used 1 and 0.54 for {\it f} and $\eta_{\rm{B}}$ (Section \ref{sec:obs}), respectively.
{\it T}$_{\rm {ex}}$ is the excitation temperature, and {\it T}$_{\rm {bg}}$ is the cosmic microwave background temperature ($\simeq2.7$ K).
{\it J}({\it T}) in Equation (\ref{tem}) is the Planck function.
In Equation (\ref{col}), {\it N} is the column density,  $\Delta v$ is the line width (FWHM), $Q$ is the partition function, $\mu$ is the permanent electric dipole moment of HC$_{3}$N ($3.73172\times10^{-18}$ esu cm, \citet{1985JChPh...82...1702}), and $E_{\rm {lower}}$ is the energy of the lower rotational energy level. 

\citet{2009ApJ...702...1025} derived the excitation temperatures and the column densities of HC$_{3}$N using the LTE analysis.
The determined excitation temperatures and column densities are $9.7\pm0.2$ K and ($2.7\pm0.2$)$\times10^{13}$ cm$^{-2}$ when the $J=5-4$ and $J=10-9$ data were included, and $16.9\pm0.5$ K and ($1.19\pm0.03$)$\times10^{13}$ cm$^{-2}$ using only the $J=10-9$ and $J=17-16$ lines.
We then derived $\tau$ assuming the excitation temperatures of 9.7 K and 16.9 K, respectively, using Equation (\ref{tau}).
The calculated column densities are summarized in Table \ref{tab:tab3}.
The column densities are determined to be ($2.61\pm0.03$)$\times10^{13}$, ($2.8\pm0.6$)$\times10^{11}$, ($3.0\pm0.5$)$\times10^{11}$, and ($3.9\pm0.6$)$\times10^{11}$ cm$^{-2}$ using the excitation temperature of 9.7 K, and ($7.87\pm0.09$)$\times10^{12}$, ($1.1\pm0.2$)$\times10^{11}$, ($1.2\pm0.2$)$\times10^{11}$, and ($1.5\pm0.2$)$\times10^{11}$ cm$^{-2}$ using the excitation temperature of 16.9K for HC$_{3}$N, H$^{13}$CCCN, HC$^{13}$CCN, and HCC$^{13}$CN, respectively.
The two assumed excitation temperatures seem to be the lower and upper limits, and then the derived column densities are also the upper and lower limits.
A change in the assumed excitation temperature by a factor of 2 does not affect the derived column densities of the three $^{13}$C isotopologues within 3-sigma errors.     
The derived column density of the normal species using the excitation temperature of 9.7 K agrees with that derived by \citet{2009ApJ...702...1025} (($2.7\pm0.2$)$\times10^{13}$ cm$^{-2}$), while the column density using the excitation temperature of 16.9 K is lower than that calculated by \citet{2009ApJ...702...1025} (($1.19\pm0.03$)$\times10^{13}$ cm$^{-2}$) by 1.5 times.
The abundance ratios of the three $^{13}$C isotopologues are derived to be 0.9 ($\pm0.2$) : 1.00 : 1.29 ($\pm0.19$) ($1\sigma$) for $[$H$^{13}$CCCN$]: [$HC$^{13}$CCN$]: [$HCC$^{13}$CN].

Using the column densities of the normal species of  ($2.7\pm0.2$)$\times10^{13}$ cm$^{-2}$ and ($1.19\pm0.03$)$\times10^{13}$ cm$^{-2}$ for the excitation temperatures of 9.7 K and 16.9 K, respectively \citep{2009ApJ...702...1025}, we also calculated the $^{12}$C/$^{13}$C ratios of HC$_{3}$N, as summarized in Table \ref{tab:tab3}.
When the excitation temperature is 9.7 K, the $^{12}$C/$^{13}$C ratios are determined to be $97\pm21$, $90\pm15$, and $70\pm10$ ($1\sigma$) for H$^{13}$CCCN, HC$^{13}$CCN, and HCC$^{13}$CN, respectively.
The $^{12}$C/$^{13}$C ratios are derived to be $108\pm23$, $102\pm18$, and $79\pm12$ ($1\sigma$) for H$^{13}$CCCN, HC$^{13}$CCN, and HCC$^{13}$CN, assuming that the excitation temperature is 16.9 K.
The $^{12}$C/$^{13}$C ratios do not change depending on the two assumed excitation temperatures within 1-sigma errors.
These ratios would be the lower and upper limits.

We also calculated the column densities of the normal species and the three $^{13}$C isotopologues of HC$_{3}$N in G28.28-0.36, using the $J=9-8$ and $J=10-9$ lines, respectively.
We conducted calculations using Equations (\ref{tau}) - (\ref{col}) with the excitation temperatures of 50, 100, and 200 K, respectively, because the typical temperature in hot cores is $\sim100$ K.
We summarize the results with the excitation temperature of 100 K in Table \ref{tab:tab4}.
The column densities of the normal species and the three $^{13}$C isotopologues are derived to be ($4.97\pm0.18$)$\times 10^{12}$, ($1.0\pm0.2$)$\times10^{11}$, ($1.04\pm0.17$)$\times10^{11}$, and ($1.53\pm0.17$)$\times10^{11}$ cm$^{-2}$ for HC$_{3}$N, H$^{13}$CCCN, HC$^{13}$CCN, and HCC$^{13}$CN, respectively, using the $J=9-8$ lines. 
Using the $J=10-9$ lines, the derived column densities of the three $^{13}$C isotopologues are ($7.1\pm1.0$)$\times10^{10}$, ($6.7\pm0.8$)$\times10^{10}$, and ($8.4\pm0.9$)$\times10^{10}$ cm$^{-2}$ for H$^{13}$CCCN, HC$^{13}$CCN, and HCC$^{13}$CN, respectively.
We discuss the differences in the column densities between the $J=9-8$ and $10-9$ lines in Section \ref{sec:ratio}.
The values of the column densities do not change depending on the assumed excitation temperatures within 1-sigma errors.
The abundance ratios are found to be 1.0 ($\pm 0.2$) : 1.00 : 1.47 ($\pm 0.17$) ($1\sigma$), and 1.05 ($\pm 0.15$) : 1.00 : 1.22 ($\pm 0.14$) ($1\sigma$) for $[$H$^{13}$CCCN$]: [$HC$^{13}$CCN$]: [$HCC$^{13}$CN] using the $J=9-8$ and $10-9$ lines, respectively.
Thus the abundance ratios are consistent between the $J=9-8$ and $10-9$ lines within 1-sigma errors.

We also derived the $^{12}$C/$^{13}$C ratios using the $J=9-8$ lines, and these results are summarized in Table \ref{tab:tab4}.
The calculated ratios are $50\pm11$, $48\pm8$, and $32\pm4$ for H$^{13}$CCCN, HC$^{13}$CCN, and HCC$^{13}$CN, respectively.
The $^{12}$C/$^{13}$C ratios do not change depending on the assumed excitation temperatures.

\floattable
\begin{deluxetable}{lcccc}
\tabletypesize{\scriptsize}
\tablecaption{Column Densities and $^{12}$C/$^{13}$C Ratios of HC$_{3}$N in L1527\label{tab:tab3}}
\tablewidth{0pt}
\tablehead{
\colhead{} & \multicolumn{2}{c}{$T_{\rm {ex}}=9.7$ K} & \multicolumn{2}{c}{$T_{\rm {ex}}=16.9$ K} \\
\cline{2-5}
\colhead{} & \colhead{Column Density} & \colhead{$^{12}$C/$^{13}$C\tablenotemark{a}} & \colhead{Column Density} & \colhead{$^{12}$C/$^{13}$C\tablenotemark{b}} \\
\colhead{Species} & \colhead{($\times 10^{11}$ cm$^{-2}$)} & \colhead{Ratio} & \colhead{($\times 10^{11}$ cm$^{-2}$)} & \colhead{Ratio}
}
\startdata
H$^{13}$CCCN & $2.8 \pm 0.6$ & $97 \pm 21$ & $1.1 \pm 0.2$ & $108 \pm 23$ \\
HC$^{13}$CCN & $3.0 \pm 0.5$ & $90 \pm 15$ & $1.2 \pm 0.2$ & $102 \pm 18$ \\
HCC$^{13}$CN & $3.9 \pm 0.6$ & $70 \pm 10$ & $1.5 \pm 0.2$ & $79 \pm 12$ \\
\enddata
\tablecomments{The error corresponds to one standard deviation.}
\tablenotetext{a}{The ratios were derived using the column density of the normal species of $2.7 \pm 0.2 \times 10^{13}$ cm$^{-2}$, which was derived by the LTE analysis using $T_{\rm {ex}} = 9.7$ K \citep{2009ApJ...702...1025}.}
\tablenotetext{b}{The ratios were derived using the column density of the normal species of $1.19 \pm 0.03 \times 10^{13}$ cm$^{-2}$, which was derived by the LTE analysis using $T_{\rm {ex}} = 16.9$ K \citep{2009ApJ...702...1025}.}
\end{deluxetable} 

\floattable
\begin{deluxetable}{lcccccc}
\tabletypesize{\scriptsize}
\tablecaption{Column Densities and $^{12}$C/$^{13}$C Ratios of HC$_{3}$N in G28.28-0.36\label{tab:tab4}}
\tablewidth{0pt}
\tablehead{
\colhead{} & \multicolumn{2}{c}{$J=9-8$} & \colhead{$J=10-9$} \\
\cline{2-4}
\colhead{} & \colhead{Column Density} & \colhead{$^{12}$C/$^{13}$C} & \colhead{Column Density} \\
\colhead{Species} & \colhead{($\times 10^{11}$ cm$^{-2}$)} & \colhead{Ratio} & \colhead{($\times 10^{10}$ cm$^{-2}$)}
}
\startdata
HC$_{3}$N  & ($4.97 \pm 0.18$) $\times 10$ & - & - \\
H$^{13}$CCCN & $1.0 \pm 0.2$ & $50 \pm 11$ & $7.1 \pm1.0$ \\
HC$^{13}$CCN & $1.04 \pm 0.17$ & $48 \pm 8$ & $6.7 \pm 0.8$ \\
HCC$^{13}$CN & $1.53 \pm 0.17$ & $32 \pm 4$ & $8.4 \pm 0.9$ \\
\enddata
\tablecomments{The error corresponds to one standard deviation. The assumed excitation temperature is 100 K.}
\end{deluxetable} 

\section{Discussion} \label{sec:dis}

\subsection{$^{13}$C Isotopic Fractionation and Formation Mechanisms of HC$_{3}$N} \label{sec:frac}

The abundance ratios of the three $^{13}$C isotopologues are derived to be 0.9 ($\pm0.2$) : 1.00 : 1.29 ($\pm0.19$) ($1\sigma$) and 1.0 ($\pm 0.2$) : 1.00 : 1.47 ($\pm 0.17$) ($1\sigma$) for $[$H$^{13}$CCCN$]: [$HC$^{13}$CCN$]: [$HCC$^{13}$CN] in L1527 and G28.28-0.36, respectively.
One possible mechanism producing $^{13}$C isotopic fractionation is isotope exchange reactions.
As \citet{1998aap...329...1156} discussed, the isotope exchange reactions can be negligible in the case of HC$_{3}$N.
Hence, the differences in the abundances among the three $^{13}$C isotopologues should occur during their formation processes.

In both L1527 and G28.28-0.36, the abundance ratios of the three $^{13}$C isotopologues of HC$_{3}$N show the following two characteristics;
\begin{enumerate}
\item The abundances of H$^{13}$CCCN and HC$^{13}$CCN are comparable with each other, and
\item The abundance of HCC$^{13}$CN is higher than the previous two species.
\end{enumerate}
These characteristics imply that the main formation pathway of HC$_{3}$N contains two equivalent carbon atoms, and the other carbon atom originates from different parent species.
We investigate possible reactions leading to HC$_{3}$N, using the UMIST Database for Astrochemistry 2012 \citep{2013A&A...550...A36}.
The four formation pathways are possible as follows.

\noindent Pathway 1: the neutral-neutral reaction between C$_{2}$H$_{2}$ and CN (Reaction (\ref{r6})),

\noindent Pathway 2: the neutral-neutral reaction between C$_{2}$H and HNC,

\noindent Pathway 3: the ion-molecule reactions between C$_{3}$H$_{n}^{+}$ ($n=3-5$) and nitrogen atoms followed by electron recombination reactions, and

\noindent Pathway 4: the ion-molecule reactions between C$_{2}$H$_{2}^{+}$ and HCN followed by electron recombination reactions.

If Pathway 1 is the main formation pathway of HC$_{3}$N, the abundance ratios of the three $^{13}$C isotopologues of HC$_{3}$N should be $1:1:x$, where $x$ is an arbitrary value, for $[$H$^{13}$CCCN$]: [$HC$^{13}$CCN$]: [$HCC$^{13}$CN], because two carbon atoms in C$_{2}$H$_{2}$ are equivalent and the triple bond between C and N in CN molecule is preserved during proceeding of the reaction \citep{1997ApJ...489...113}.
Our observational results show 0.9 ($\pm0.2$) : 1.00 : 1.29 ($\pm0.19$) ($1\sigma$) and 1.0 ($\pm 0.2$) : 1.0 : 1.47 ($\pm 0.17$) ($1\sigma$) for $[$H$^{13}$CCCN$]: [$HC$^{13}$CCN$]: [$HCC$^{13}$CN] in L1527 and G28.28-0.36, respectively.
Therefore, our observational results are well consistent with the expected ratios by Pathway 1.

In L1527, \citet{2008ApJ...681...1385} showed the formation pathways leading to C$_{2}$H$_{2}$ from CH$_{4}$ as follows;

\begin{equation} \label{r1}
{\mathrm C}^{+} + {\mathrm {CH}}_{4} \rightarrow {\mathrm C}_{2}{\mathrm H}_{3}^{+} + {\mathrm H},
\end{equation}

\begin{equation} \label{r2}
{\mathrm C}^{+} + {\mathrm {CH}}_{4} \rightarrow {\mathrm C}_{2}{\mathrm H}_{2}^{+} + {\mathrm H}_{2},
\end{equation}
followed by

\begin{equation}\label{r3}
{\mathrm C}_{2}{\mathrm H}_{2}^{+} + {\mathrm H}_{2} \rightarrow {\mathrm C}_{2}{\mathrm H}_{4}^{+},
\end{equation}
and

\begin{equation} \label{r4}
{\mathrm C}_{2}{\mathrm H}_{3}^{+} + {\mathrm e} \rightarrow {\mathrm C}_{2}{\mathrm H}_{2} + {\mathrm H},
\end{equation}

\begin{equation} \label{r5}
{\mathrm C}_{2}{\mathrm H}_{4}^{+} + {\mathrm e} \rightarrow {\mathrm C}_{2}{\mathrm H}_{2} + 2{\mathrm H}.
\end{equation}
C$_{2}$H$_{2}$ then can be efficiently formed from CH$_{4}$, and the neutral-neutral reaction between C$_{2}$H$_{2}$ and CN seems to occur.  
The chemical model calculation supports our conclusion derived from our observations.
\citet{2009MNRAS...394...221} showed that the neutral-neutral reaction between C$_{2}$H$_{2}$ and CN (Reaction (\ref{r6})) proceeds under the hot core condition, and C$_{2}$H$_{2}$ is released from the grain mantle inside hot cores.
Our conclusion about G28.28-0.36 is consistent with their chemical model calculation.
The validity of their model calculation seems to be supported by detections of HC$_{5}$N \citep{2014MNRAS...443...2252} and HC$_{7}$N (Taniguchi et al., {\it {in prep.}}).

We next consider a possibility that Pathway 2 is the dominant formation reaction leading to HC$_{3}$N.
In this reaction, three carbon atoms are not equivalent, and the expected abundance ratios are $x:y:z$ for $[$H$^{13}$CCCN$]: [$HC$^{13}$CCN$]: [$HCC$^{13}$CN], where $x$, $y$, and $z$ are arbitrary values.
In fact, \citet{2010A&A...512...A31} showed that the [C$^{13}$CH]/[$^{13}$CCH] abundance ratio is $1.6\pm0.1$ ($3\sigma$).
\citet{1997ApJ...489...113} also demonstrated their quantum chemical calculations, and their results show that a carbon atom in HNC and a carbon atom with an unpaired electron in CCH are connected.
From the above two results, the abundance ratios of HC$_{3}$N should be $[$H$^{13}$CCCN$]: [$HC$^{13}$CCN$]: [$HCC$^{13}$CN] = $1.6 : 1.0 : x$, where $x$ is an arbitrary value, if Pathway 2 is the main formation pathway of HC$_{3}$N.
Pathway 2 cannot explain our observational results, and we conclude that the reaction between C$_{2}$H and HNC is not the primary formation pathway of HC$_{3}$N.

Pathways 3 and 4 include the ion-molecule reactions.
There is a possibility that there is no significant differences in abundances among all of the $^{13}$C isotopologus, when the main formation mechanism is the ion-molecule reactions \citep{2016ApJ...817...147}.
We also consider that scrambling may occur during the processes of these ion-molecule reactions.
The clear differences in abundances among the three $^{13}$C isotopologues of HC$_{3}$N should not be recognized, if scrambling occurs.
Hence, these ion-molecule reactions are not the main formation mechanisms of HC$_{3}$N in L1527 and G28.28-0.36.

In summary, the abundance ratios derived by our observations agree with the ratios of only Pathway 1.
We thus propose that the neutral-neutral reaction between C$_{2}$H$_{2}$ and CN (Reaction (\ref{r6})) dominates other formation pathways in both L1527 and G28.28-0.36.
This proposal also agrees with the predicted primary formation pathway of HC$_{3}$N by chemical model calculations \citep{2008ApJ...681...1385,2009MNRAS...394...221}.

\subsection{$^{12}$C/$^{13}$C Ratios in L1527 and G28.28-0.36} \label{sec:ratio}

The $^{12}$C/$^{13}$C ratios of HC$_{3}$N in L1527 are summarized in Table \ref{tab:tab3}.
When we assume that the excitation temperature is 9.7 K, the $^{12}$C/$^{13}$C ratios are consistent with the elemental ratio of $60-70$ in the local interstellar medium (ISM) \citep{1993ApJ...408...539,2002ApJ...578...211,2005ApJ...634...1126} within 1-sigma errors.
The derived $^{12}$C/$^{13}$C ratios of H$^{13}$CCCN and HC$^{13}$CCN are slightly higher than the elemental ratio in the local ISM, using the excitation temperature of 16.9 K.
However, the ratios of these two $^{13}$C isotopologues are consistent with the elemental ratio in the local ISM within 2-sigma errors.
The $^{12}$C/$^{13}$C ratio of HCC$^{13}$CN agrees with the elemental ratio of the local ISM within 1-sigma error.
Therefore, the $^{13}$C species of HC$_{3}$N are not significantly diluted in L1527, as well as TMC-1 CP \citep{1998aap...329...1156}.

In Table \ref{tab:tab4}, the $^{12}$C/$^{13}$C ratios of the three $^{13}$C isotopologues in G28.28-0.36 are summarized.
These $^{12}$C/$^{13}$C values are lower than the elemental ratio of $60-70$ in the local ISM, because G28.28-0.36 is located nearer the Galactic center than the Earth and L1527.
The $^{12}$C/$^{13}$C ratio shows a gradient with Galactic distance ($D_{\rm {GC}}$) \citep{2002ApJ...578...211,2005ApJ...634...1126}.
We estimated the $D_{\rm {GC}}$ of G28.28-0.36 to be 5.4 kpc using the trigonometry.
The $^{12}$C/$^{13}$C ratios at $D_{\rm {GC}}=5.4$ kpc are derived to be $39-65$, using the results obtained by the observations of CN, CO, and H$_{2}$CO ($^{12}$C/$^{13}$C = 6.21(1.00)$D_{\rm {GC}}$+18.71(7.37)) \citep{2005ApJ...634...1126}.
The $^{12}$C/$^{13}$C ratios of HC$_{3}$N obtained by our observations are well consistent with the estimated values at $D_{\rm {GC}}=5.4$ kpc.
These suggest that $^{13}$C isotopologues of HC$_{3}$N are not heavily diluted, which is in good agreement with in the local ISM.

The column densities derived by the $J=10-9$ lines are lower than those derived by the $J=9-8$ lines by $1.3-1.9$ times.
The explanation for the differences in the column densities between the $J=10-9$ lines and $J=9-8$ lines is that the spatial distribution of HC$_{3}$N shows the ring-like structure with outer strong emission peaks, which is supported by our high spatial resolution map with the Very Large Array (Taniguchi et al., {\it {in prep.}}).

\subsection{Comparison of $^{13}$C Isotopic Fractionation in Various Sources}

\floattable
\begin{deluxetable}{cccccccc}
\tabletypesize{\scriptsize}
\tablecaption{$^{13}$C Isotopic Fractionation of HC$_{3}$N in Various Sources\label{tab:tab5}}
\tablehead{
\colhead{Source} & \colhead{H$^{13}$CCCN} & \colhead{HC$^{13}$CCN} & \colhead{HCC$^{13}$CN} & source type & Temperature (K) & Density (cm$^{-3}$)
}
\startdata
L1527 & 0.9 ($\pm 0.2$) & 1.00 & 1.29 ($\pm 0.19$) & WCCC & $20-30$ & $\sim10^{5}$\tablenotemark{c} \\
G28.28-0.36 & 1.0 ($\pm0.2$) & 1.00 & 1.47 ($\pm0.17$) & hot core & $100-200$ & $\sim10^{6}$ \\
TMC-1 CP\tablenotemark{a} & 1.0 & 1.0 & 1.4 ($\pm0.2$) & dark cloud & $\sim10$ & $\sim10^{4}$ \\
Serpens South 1A\tablenotemark{b} & 0.91 ($\pm 0.09$) & 1.00 & 1.32 ($\pm 0.09$) & IRDC & $\sim15$ & $\sim10^{5}$ \\
\enddata
\tablecomments{The error corresponds to one standard deviation.}
\tablenotetext{a}{The ratios were derived using the column densities \citep{1998aap...329...1156}.}
\tablenotetext{b}{The ratios were derived using the column densities \citep{2016...ApJ...824...136}.}
\tablenotetext{c}{\citet{2001pasj...53...251}.}

\end{deluxetable}

We summarize $^{13}$C isotopic fractionation of HC$_{3}$N in the four sources, using HC$^{13}$CCN as a reference, in Table \ref{tab:tab5}.
We also summarize the source types and their typical temperatures and densities in Table \ref{tab:tab5}.
We categorize these four sources into the star-forming cores (L1527 and G28.28-0.36) and the starless cores (TMC-1 CP and Serpens South 1A).
We can also categorize these four sources into the low-mass star forming regions (TMC-1 CP and L1527) and the high-mass star forming regions (Serpens South 1A and G28.28-0.36).
We then compare various environment in star forming regions.
Although there are wide ranges of temperatures ($\sim10\; {\rm {K}}-200\; {\rm {K}}$) and densities ($10^{4}\; {\rm {cm}}^{-3}-10^{6}\; {\rm {cm}}^{-3}$), the $^{13}$C isotopic fractionation pattern in all of the four sources show the same tendency.
The abundances of H$^{13}$CCCN and HC$^{13}$CCN are comparable with each other, and HCC$^{13}$CN is more abundant than the others.
In addition, the [HCC$^{13}$CN]/[HC$^{13}$CCN] ratios are in good agreement ($\sim1.3$) among the four sources within 1-sigma errors.
These results may imply that the neutral-neutral reaction between C$_{2}$H$_{2}$ and CN (Reaction (\ref{r6})) is the universal main formation mechanism of HC$_{3}$N.

\section{Conclusions} \label{sec:con}

We carried out observations of the $J=9-8$ and $10-9$ rotational transitions of the three $^{13}$C isotopologues of HC$_{3}$N toward the low-mass star forming region L1527 and the high-mass star forming region G28.28-0.36 with the NRO 45-m telescope.
The abundance ratios are found to be 0.9 ($\pm0.2$) : 1.00 : 1.29 ($\pm0.19$) ($1\sigma$) and 1.0 ($\pm 0.2$) : 1.00 : 1.47 ($\pm 0.17$) ($1\sigma$) for $[$H$^{13}$CCCN$]: [$HC$^{13}$CCN$]: [$HCC$^{13}$CN] in L1527 and G28.28-0.36, respectively.
Our observational results suggest that the neutral-neutral reaction between C$_{2}$H$_{2}$ and CN seem to overwhelm the other formation pathways in both L1527 and G28.28-0.36. 
In addition, the $^{13}$C isotopic fractionation pattern seen in the two star-forming regions is the same one in starless cores.
The primary formation pathway of HC$_{3}$N may be common from low-mass prestellar cores to high-mass star-forming cores.



\acknowledgments

We would like to express our great thanks to the staff of the Nobeyama Radio Observatory.
The Nobeyama Radio Observatory is a branch of the National Astronomical Observatory of Japan, National Institutes of Natural Sciences.
In particular, we deeply appreciate Dr. Tetsuhiro Minamidani, Ms. Chieko Miyazawa, and Dr. Hiroyuki Nishitani for immediate responding to the equipment problems.
We express our thanks to Mr. Mitsuhiro Matsuo (Ph. D student of Kagoshima University) for helping observations at Nobeyama.
This work was supported in part by the Center for the Promotion of Integrated Sciences (CPIS) of SOKENDAI.

\listofchanges

\end{document}